# Cryptographic Strengthening of MST3 cryptosystem via Automorphism Group of Suzuki Function Fields

Gennady Khalimov[1[0000-0002-2054-9186]], Yevgen Kotukh [2[0000-0003-4997-620X]]

[1]Kharkiv National University of Radioelectronics, Kharkiv, 61166, Ukraine hennadii.khalimov@nure.ua
[2]Yevhenii Bereznyak Military Academy, Kyiv, Ukraine yevgenkotukh@gmail.com

**Abstract.** The article describes a new implementation of MST3 cryptosystems based on the automorphism group of the field of the Suzuki function. The main difference in the presented implementation is to use the logarithmic signature for encryption not only in the center of the group, as in the well-known implementation of MST3 for Suzuki groups but also for coordinates outside the center of the group. The presented implementation of a cryptosystem has greater reliability. The complexity of cryptanalysis and the size of the message for encryption squared is greater than that of the MST3 cryptosystem in the Suzuki group.

## INTRODUCTION

The rapid development of quantum computing poses a significant threat to classical cryptographic systems, particularly those relying on the computational difficulty of integer factorization and discrete logarithm problems. Many widely used public key cryptosystems are expected to become vulnerable in the quantum era. Therefore, the urgent task of designing cryptosystems that can withstand quantum attacks has become a key focus of contemporary cryptographic research.

In the early 2000s, Shpilrain and Ushakov proposed cryptographic constructions based on computationally unsolvable problems within algebraic structures [1]. This idea sparked the development of various group-based cryptosystems [2–5], with earlier roots traceable to the work of Wagner and Magyarik [6], who utilized permutation groups for public key cryptography.

One significant advancement in this area came from Magliveras [4], who introduced the concept of logarithmic signatures for finite permutation groups. This approach draws upon positional number representations using polybasic bases, with group operations applied to the expansion coefficients. Lempken et al. [7] further extended this idea through the notion of random covers, proposing cryptographic schemes wherein private keys are composed of manually selected logarithmic signatures and public keys consist of arrays of random transformations.

Further refinements were made by Svaba and van Trung [8], who introduced matrix transformations into the random covering framework. More recently, an MST3 implementation using the automorphism group of the Hermitian function field demonstrated that encryption schemes based on large-order groups and advanced logarithmic signatures can achieve higher levels of security [9]. It was shown here that it is possible to construct encryption schemes based on logarithmic signatures of higher secrecy on a large-order group. In this paper, we present an encryption scheme that employs the automorphism group of the Suzuki function field — a group of even larger order than both the Suzuki group and the Hermitian function field automorphism group. This enlargement yields notable security benefits and greater implementation strength in cryptographic applications.

## THE AUTOMORPHISM GROUP OF THE SUZUKI FUNCTION FIELD

Let $F_q$ be a finite field and $F/F_q$ be an algebraic function field over the full constant field $F_q$ of genus $g$. The Suzuki function field is an optimal function field defined over the finite field with an even characteristic $p$. Let $p=2$, $q=2^n$ with $n=2s+1$ where $s \in N\setminus\{0\}$, and $q_0=2^s$. Let $K$ be the finite field $F_q$.

The Suzuki function field over $K$ is defined as $S=K(x,y)$ where $y^q+y=x^{2q_0}(x^q+x)$. $S/K(x)$ is a Galois extension of the degree $q$, and the pole of $x$ is totally ramified in the extension $S/K(x)$. Let $P_\infty$ denote the unique rational place of $S$ lying over the pole of $x$. The genus of $S$ is $g(S)=q_0(q-1)$, and the number of rational places of $S$ is $q^2+1$.

The automorphism group of $S$ denoted by $A=Aut(S/K)$ is isomorphic to the Suzuki group $Sz(q)$ which has order $ordA=\left(q^2+1\right)q^2\left(q-1\right)$.

The decomposition group of $P_\infty$ which is denoted as $A(P_\infty)$ consists of the automorphisms of $S|F_q$ acting on it as

$$\begin{cases} \sigma(x) = ax + b \\ \sigma(y) = a^{2q_0+1}y + ab^{2q_0}x + c, \end{cases}$$

where $a \in F_q^* := F_q \setminus \{0\}$, $b,c \in F_q$ [10].

An involution $\psi \in A$ is given by

$$\begin{cases} \psi(x) = y/h \\ \psi(y) = y/h \end{cases}$$

where $h = xy + x^{2q_0+2} + y^{2q_0}$.

The automorphism group of $S$ is generated by $A(P_\infty)$ and $\psi$, that is, $Sz(q) \cong \langle A(P_\infty), \psi \rangle$.

Once again, an automorphism in the decomposition group $A(P_\infty)$ can be identified with a triple $[a,b,c]$. More precisely, $A(P_\infty) = [a,b,c]$, $a,b,c \in K$, and $a = 0$. The group structure is given by $[a_1,b_1,c_1]\cdot[a_2,b_2,c_2] = \left[a_1a_2, a_2b_1 + b_2, a_2^{2q_0+1}c_1 + a_2b_2^{2q_0}b_1 + c_2\right]$.

The identity is the triple $[1,0,0]$ and the inverse of $[a,b,c]$ is

$$[a,b,c]^{-1} = \left[a^{-1}, -a^{-1}b, \left(a^{-1}b\right)^{2q_0+1} - a^{-(2q_0+1)}c\right].$$

It has order $ordA(P_\infty) = q^2(q-1)$.

**Remark 1.** The order group $A(P_\infty)$ is greater than Suzuki group. Suzuki group is isomorphic to the projective linear group $PGL(3, F_q)$, where $q = 2q_0^2$, $q_0 = 2^n$ and has order $q^2$ and included in MST3 cryptosystems. A larger group order gives an advantage to cryptosystem secrecy.

## MST CRYPTOSYSTEMS

The basic idea of MST cryptosystems is to surjective mapping input message into an element of group using a conversion key. Formalization of computations is given by the following definition [11].

**Definition 1**. Let $\alpha = [A_1, \ldots, A_s]$ be a logarithmic signature of type $(r_1, r_2, \ldots, r_s)$ for $G$ with $A_i = \left[a_{i,1}, a_{i,2}, \ldots, a_{i,r_i}\right]$, where $m = \prod_{i=1}^{s} r_i$. Let $m_1 = 1$ and $m_i = \prod_{j=1}^{i-1} r_j$ for $i = 2, \ldots, s$. Let $\tau$ denote the canonical bijection $\tau : \mathbb{Z}_{r_1} \times \mathbb{Z}_{r_2} \times \ldots \times \mathbb{Z}_{r_s} \to \mathbb{Z}_m$, $\tau(j_1, j_2, \ldots, j_s) = \sum_{i=1}^{s} j_i \cdot m_i$.

Then the surjective mapping $\alpha' : \mathbb{Z}_m \to G$ induced by is $\alpha'(x) = a_{1j_1} \cdot a_{2j_2} \cdots a_{sj_s}$ where $(j_1, j_2, \ldots, j_s) = \tau^{-1}(x)$.

More generally, if $\alpha = [A_1, \ldots, A_s]$ is a logarithmic signature, then each element $g \in G$ can be expressed uniquely (at least one way) as a product of the form $g = a_1 \cdot a_2 \cdots a_s$, for $a_i \in A_i$ [7].

Let $G$ is an ultimate nonabelian group with a nontrivial center $\mathbb{Z}$, such that $G$ does not decompose over $\mathbb{Z}$. Suppose that $\mathbb{Z}$ is quite large, such that the search is over $\mathbb{Z}$ is computationally impracticable.

The cryptographic hypothesis, which is the basis for the cryptosystem, is that if $\alpha = [\mathrm{A}_1, \mathrm{A}_2, \ldots, \mathrm{A}_s] := (a_{i,j})$ – accidental cover for a "large" matrices $S$ at $G$, then search for the layout $g = a_{1j_1} a_{2j_2} \cdots a_{sj_s}$ for any element $g \in G$ relatively $\alpha$ is, in general, not a solvable problem. There are several encryption algorithms for MST cryptosystems. The latest version known as MST3 has an implementation for the Suzuki group.

**THE MAIN STEPS OF THE ENCRYPTION ALGORITHM.**

The following steps from 1 to 5 introduce the calculation of public and secret keys.

Step 1. Generating tame logarithmic signatures $\beta = [B_1, B_2, \ldots, B_s] := (\mathrm{b}_{ij})$ the class $(r_1, r_2, \ldots, r_s)$ for $\mathbb{Z}$

Step 2. Generating a random cover $\alpha = [\mathrm{A}_1, \mathrm{A}_2, \ldots, \mathrm{A}_s] := (a_{i,j})$ the same class as i $\beta$ for some subset $J$ from $G$ such that $\mathrm{A}_1, \ldots, \mathrm{A}_s \subseteq G \setminus \mathbb{Z}$;

Step 3. Generating a set of elements $t_0, t_1 \ldots, t_s \in G \setminus \mathbb{Z}$;

Step 4. Define homomorphism to calculate $f : G \to \mathbb{Z}$;

Step 5. Calculating $\gamma := (h_{ij}) = (t_{i-1}^{-1} f(a_{ij}) b_{ij} t_i)$ for $i = 1,...,s$, $j = 1,...,r_i$.

As a result, we get the public key $(\alpha = (a_{ij}), \gamma = (h_{ij}), f)$ and private key $(\beta = (\mathrm{b}_{ij}), (\mathrm{t}_0, ...., \mathrm{t}_s))$.

Steps 6-8 represent an encryption stage as follows:

Step 6. Set a random number $R \in Z_{|\mathbb{Z}|}$. Let the message to be encrypted $x \in \mathbb{Z}_{|Z|}$.

Step 7. Calculate $y_1 = \alpha'(R) \cdot x$, $y_2 = \gamma(R) = t_0^{-1} f(\alpha'(R)) b'(R) t_s$.

Step 8. Transmit $y = (y_1, y_2)$.

Steps 6-8 represent a decryption stage as follows:

For decryption, we have the cipher text $y = (y_1, y_2)$, private key $(\beta = (\mathrm{b}_{ij}), (\mathrm{t}_0, ...., \mathrm{t}_s))$ and the function of the homomorphism $f : G \to \mathbb{Z}$.

Step 9. Let's calculate $\beta(R) = y_2 t_s^{-1} f(y_1)^{-1} t_0$.

Step 10. Checking is determined by the fact that

$$y_2 = \gamma(R) = t_0^{-1} f(a_{1j_1}(R)) b_{1j_1}(R) t_1 \cdots t_{s-1}^{-1} f(a_{sj_s}(R)) b_{sj_s}(R) t_s$$
$$= b_{1j_1}(R) t_0^{-1} f(a_{1j_1}(R)) t_1 \cdots b_{sj_s}(R) t_{s-1}^{-1} f(a_{sj_s}(R)) t_s = \beta(R) t_0^{-1} f(\alpha(R)) t_s = \beta(R) t_0^{-1} f(y_1) t_s$$

Step 11. Recover $R$ with $\beta(R)$ using $\beta^{-1}$, because $\beta$ is simple.

Step 12. Calculate $x = \alpha'(R)^{-1} \cdot y_1$.

**Remark 2.** Message $x$ masked by a logarithmic signature on the arrays $\alpha = (a_{ij})$, $\gamma = (h_{ij})$ which is calculated for a random number $R$. The function of the homomorphism $f : G \to \mathbb{Z}$ moves the group element to the center of the group. Calculation $\beta(R) = y_2 t_s^{-1} f(y_1)^{-1} t_0$ and the subsequent recovery of $R$ is possible due to the commutativity of the center.

## ENCRYPTION SCHEME BASED ON THE AUTOMORPHISM GROUP OF THE SUZUKI FUNCTION FIELD

Our suggestion is to use the logarithmic signature for encryption not only in the center of group $A(P_\infty)$, as in the well-known implementation of MST3 for Suzuki groups, but also for other coordinates outside the center of the group.

Remark 1 implies that for the construction of the MST3 cryptosystem, the advantage is given to the group $A(P_\infty)$ based on the automorphism $\sigma(x),\sigma(y)$.

Each element of $A(P_\infty)$ can be expressed uniquely

$$A(P_\infty)=\left\{S(a,b,c)\middle|a\in F_q^*:=F_q\setminus\{0\},c,b\in F_q\right\}$$

where $S(a,b,c)=[a,b,c]$ and group operation is defined as

$[a_1,b_1,c_1]\cdot[a_2,b_2,c_2]=\left[a_1a_2,a_2b_1+b_2,a_2^{2q_0+1}c_1+a_2b_2^{2q_0}b_1+c_2\right]$ and the inverse of $S(a,b,c)$ is

$S(a,b,c)^{-1}=S\left(a^{-1},-a^{-1}b,\left(a^{-1}b\right)^{2q_0+1}-a^{-(2q_0+1)}c\right)$.

It is simple to show by direct calculations. The identity is the triple $S(1,0,0)$. It follows that $|A(P_\infty)|=q^2(q-1)$. The center $Z\left(A(P_\infty)\right)=\left\{S(1,0,c)\middle|c\in F_q\right\}$ and $\left|Z\left(A(P_\infty)\right)\right|=q$.

Next, we introduce the main steps of our encryption scheme.

## KEY GENERATION STAGE

We have a large group on the field $F_q$, $q=2q_0^{\,2}$, $q_0=2^s$ and

$A(P_\infty)=\left\{S(a,b,c)\middle|a\in F_q^*:=F_q\setminus\{0\},c,b\in F_q\right\}$ as input data for key generation stage. We expect to have a public key $[\alpha,\gamma,f]$ with the corresponding private key $[\beta,(t_0,...,t_s)]$ as output data. Here are the steps to follow:

Step 1. Choose a first tame logarithmic signature $\beta_{(1)}=\left[B_{1(1)},...,B_{s(1)}\right]=\left(b_{ij}\right)_{(1)}=S\left(1,b_{ij(1)},0\right)$ of type $\left(r_{1(1)},...,r_{s(1)}\right)$, $i=\overline{1,s(1)}$, $j=\overline{1,r_{i(1)}}$, $b_{ij(1)}\in F_q$.

Step 2. Choose a second tame logarithmic signature $\beta_{(2)}=\left[B_{1(2)},...,B_{s(2)}\right]=\left(b_{ij}\right)_{(2)}=S\left(1,0,b_{ij(2)}\right)$ of type $\left(r_{1(2)},...,r_{s(2)}\right)$, $i=\overline{1,s(2)}$, $j=\overline{1,r_{i(2)}}$, $b_{ij(2)}\in F_q$.

Step 3. Select a first random cover $\alpha_{(1)}=\left[A_{1(1)},...,A_{s(1)}\right]=\left(a_{ij}\right)_{(1)}=S\left(a_{ij(1)_1},a_{ij(1)_2},a_{ij(1)_3}\right)$ of the same type as $\beta_{(1)}$, where $a_{ij}\in A(P_\infty)$, $a_{ij(1)_1},a_{ij(1)_2},a_{ij(1)_3}\in F_q\setminus\{0\}$.

Step 4. Select a second random cover $\alpha_{(2)}=\left[A_{1(2)},...,A_{s(2)}\right]=\left(a_{ij}\right)_{(2)}=S\left(a_{ij(2)_1},a_{ij(2)_2},a_{ij(2)_3}\right)$ of the same type as $\beta_{(2)}$, where $a_{ij(2)_1},a_{ij(2)_2},a_{ij(2)_3}\in F_q\setminus\{0\}$.

Step 5. Choose $t_{0(k)},t_{1(k)},...,t_{s(k)} \in A(P_\infty)\setminus Z$, $t_{i(k)} = S\left(t_{i(k)_1},t_{i(k)_2},t_{i(k)_3}\right)$, $t_{i(k)_j} \in F^\times$, $i=\overline{0,s(k)}$, $j=\overline{1,2}$, $k=\overline{1,2}$. Let's $t_{s(1)} = t_{0(2)}$.

Step 6. Construct a homomorphism $f_1$ defined by $f_1(S(a_1,a_2,a_3)) = S(1,a_1,a_2)$.

Step 7. Calculating $\gamma_{(1)} = \left[h_{1(1)},...,h_{s(1)}\right] = \left(h_{ij}\right)_{(1)} = t^{-1}_{(i-1)(1)} f_1\left(\left(a_{ij}\right)_{(1)}\right)\left(b_{ij}\right)_{(1)} t_{i(1)}$, $i=\overline{1,s(1)}$, $j=\overline{1,r_{i(1)}}$

where $f_1\left(\left(a_{ij}\right)_{(1)}\right)\left(b_{ij}\right)_{(1)} = S\left(1, a_{ij(1)_1}+b_{ij(1)}, a_{ij(1)_2}+a_{ij(1)_1}b^{q_0}_{ij(1)}\right)$,

Step 8. Define a homomorphism $f_2$ - $f_2(S(1,a_2,a_3)) = S(1,0,a_2)$.

Step 9. Calculating $\gamma_{(2)} = \left[h_{1(2)},...,h_{s(2)}\right] = \left(h_{ij}\right)_{(2)} = t^{-1}_{(i-1)(2)} f_2\left(\left(a_{ij}\right)_{(2)}\right)\left(b_{ij}\right)_{(2)} t_{i(2)}$,

$i=\overline{1,s(2)}$, $j=\overline{1,r_{i(2)}}$ where $f_2\left(\left(a_{ij}\right)_{(2)}\right)\left(b_{ij}\right)_{(2)} = S\left(1,0,a_{ij(2)_2}+b_{ij(2)}\right)$.

An output public key $[f_1, f_2, (\alpha_k,\gamma_k)]$, and a private key $\left[\beta_{(k)},\left(t_{0(k)},...,t_{s(k)}\right)\right]$, $k=\overline{1,2}$.

## ENCRYPTION STAGE

We have a message $m \in A(P_\infty)$, $m = S(m_1,m_2,m_3)$, $m_1 \in F_q \setminus \{0\}$, $m_2,m_3 \in F_q$ and the public key $[f_1, f_2, (\alpha_k,\gamma_k)]$, $k=\overline{1,2}$ as an input data for encryption stage. We expect to have ciphertext $(y_1,y_2,y_3,y_4)$ of the message $m$ as an output.

Step 10. Choose a random $R=(R_1,R_2)$, $R_1,R_2 \in Z_{|F_q|}$.

Step 11. Calculating

$$\begin{aligned}&y_1 = \alpha'(R)\cdot m = \alpha_1'(R_1)\cdot\alpha_2'(R_2)\cdot m = S\left(a_{(1)_1}(R_1)a_{(2)_1}(R_2), a_{(2)_1}(R_2)a_{(1)_2}(R_1)+a_{(2)_2}(R_2),\right.\\&a_{(2)_1}(R_2)^{2q_0+1}a_{(1)_3}(R_1)+a_{(2)_1}(R_2)a_{(2)_2}(R_2)^{2q_0}a_{(1)_2}(R_1)+a_{(2)_3}(R_2)\ )\cdot m = S(\ a_{(1)_1}(R_1)a_{(2)_1}(R_2)m_1,\\&(a_{(2)_1}(R_2)a_{(1)_2}(R_1)+a_{(2)_2}(R_2))m_1+m_2, m_3+*)\end{aligned}$$

Here, the $(*)$ components are determined by cross-calculations in the group operation of the product of $a_{(i)_j}(R_i)$.

Step 12. Calculating

$$y_2 = \gamma'(R) = \gamma_1'(R_1)\cdot\gamma_2'(R_2) = S\left(*, a_{(1)_1}(R_1)+\beta_{(1)}(R_1)+*, a_{(2)_2}(R_2)+\beta_{(2)}(R_2)+*\right).$$

Here, the $(*)$ components are determined by cross-calculations in the group operation of the product of $t_{0(k)},...,t_{s(k)}$ and for third coordinate is added the product of $a_{(1)_1}(R_1)+\beta_{(1)}(R_1)$.

Step 13. Calculating $y_3 = f_1(\alpha_1'(R_1)) = S(1, a_{(1)_1}(R_1), a_{(1)_1}(R_1)+a_{(1)_2}(R_1)+*)$ and $y_4 = f_2(\alpha_2'(R_2)) = S(1,0,a_{(2)_2}(R_2))$. Here, the $(*)$ components are determined by cross-calculations in the group operation of the product of $a_{(1)_1}(R_1)$.

So, we have got a ciphertext $(y_1, y_2, y_3, y_4)$.

## DECRYPTION STAGE

We have a ciphertext $(y_1, y_2, y_3, y_4)$ and private key $\left[\beta_{(k)}, \left(t_{0(k)},...,t_{s(k)}\right)\right]$, $k=\overline{1,2}$ as an input data for decryption stage. We expect to receive the message $m \in A(P_\infty)$ corresponding to ciphertext $(y_1, y_2, y_3, y_4)$. To decrypt a message $m$, we need to restore random numbers $R=(R_1, R_2)$. The parameter $a_{(1)_1}(R_1)$ is known from the $y_3$ and it is included in the second component of $y_2$.

Step 14. Calculating $D^{(1)}(R_1,R_2) = t_{0(1)} \cdot y_2 t_{s(2)}^{-1} = S\left(1, a_{(1)_1}(R_1)+\beta_{(1)}(R_1), a_{(2)_2}(R_2)+\beta_{(2)}(R_2)+*\right)$ and $D^*(R) = y_3^{-1} D^{(1)}(R_1,R_2) = S\left(1, \beta_{(1)}(R_1), a_{(2)_2}(R_2)+\beta_{(2)}(R_2)+*\right)$.

Step 15. Restore $R_1$ with $\beta_{(1)}(R_1)$ using $\beta_{(1)}(R_1)^{-1}$, because $\beta$ is simple. For further calculation, it is necessary to remove the component of the array $\gamma_1'(R_1)$ from $y_2$.

Step 16. Calculating $y_2^{(1)} = \gamma_1'(R_1)^{-1} y_2 = \gamma_2'(R_2) = S\left(*,*,a_{(2)_2}(R_2)+\beta_{(2)}(R_2)+*\right)$.

Step 17. Repeat the calculations for restore $R_1$ of the $y_2^{(1)}$

$D^{(2)}(R_2) = t_{0(2)} \cdot y_2^{(1)} t_{s(2)}^{-1} = S\left(1,0,a_{(2)_2}(R_2)+\beta_{(2)}(R_2)\right)$ and $D^*(R) = y_4^{-1} D^{(2)}(R_2) = S\left(1,0,\beta_{(2)}(R_2)\right)$.

Step 18. Restore $R_2$ with $\beta_{(2)}(R_2)$ using $\beta_{(2)}(R_2)^{-1}$.

We obtain the recovery of $R=(R_1,R_2)$ and the message $m$ from $y_1$ - $m = \alpha'(R_1,R_2)^{-1} \cdot y_1$

.

## SECURITY ANALYSIS

We consider the following types of attacks.

Attack 1. Brute force attack on cipher text.

The main attack mechanics consider that by selecting $R=(R_1,R_2)$ attacker will try to decipher the text $y_1=\alpha'(R)\cdot m=\alpha_1'(R_1)\cdot\alpha_2'(R_2)\cdot m$. Difficulty of attack equal to $q^2$.

Attack 2. Brute force attack on $R=(R_1,R_2)$

The main attack mechanics consider that by selecting $R=(R_1,R_2)$ attacker will try to match $y_2=\gamma'(R)=\gamma_1'(R_1)\cdot\gamma_2'(R_2)$. Difficulty of attack equal to $q^2$.

Attack 3. Attack on session key vector.

Attacker to choose $R_1$ to match the value of $a_{(1)_1}(R_1)$ in the vector $y_1$ and choose $R_2$ to match the value of $a_{(2)_2}(R_2)$ in the vector $y_2$. Difficulty of attack equal to $q$. Possible protection mechanism - link $R_1$ and $R_2$ through matrix transformation.

Attack 4. Attacker to brute force attack on $\left(t_{0(k)},\ldots,t_{s(k)}\right)$.

Difficulty of attack equal to $q^3$.

Attack 5. Attack on the algorithm itself.

Extraction parameters $a_{(1)_1}(R_1)$, $a_{(2)_2}(R_2)$ of $y_3$, $y_4$ does not allow to calculate $\alpha_1'(R_1)\cdot\alpha_2'(R_2)$ in $y_1=\alpha_1'(R_1)\cdot\alpha_2'(R_2)\cdot m$. A simple search of parameters $R_1,R_2$ leads to brute force attack with complexity $q^2$. Since the automorphism group $A(P_\infty)$ of the Suzuki function field is defined over a large field $F_q$, the attack is not computationally possible.

## CONCLUSION

The proposed solution improves existing ones with the following results. First, the encryption scheme based on the automorphism group $A(P_\infty)$ of the Suzuki function field over $F_q$ has a higher secrecy which equals to ~ $q^2$. Second, the length of ciphertext is $3\log q$ for computing in the finite field over $F_q$ . Third, computing in the finite field is more efficient by comparison with Suzuki group cryptosystems. Also, the encryption on a finite field is three times faster in dimension compared to a cryptosystem based on Suzuki groups. Finally, the length of the logarithmic signature array is determined by the finite field over $F_q$ and significantly less compared to the Suzuki cryptosystem.

The implementation of a cryptosystem on the automorphism group $A(P_\infty)$ of the Suzuki function field requires the construction of a logarithmic signature $\beta$ on vectors $2^h$, where $h$ is determined by the size of type $r_i = 2^h$. Note that all blocks $B_i$ are subgroups of the $U(q) = \{S(1,b,c) | b,c \in F_q\}$. The size of the arrays $\beta$ and α is determined by the type $(r_1,\ldots,r_s)_b$ and $(r_1,\ldots,r_s)_c$ for coordinate $b,c$ for the subgroups $U(q)$. For 128-bit cryptography, which is equivalent to calculations over field $q = 2^{64}$, if the $r_i$ type is $r_i = 2^2$, $s = 32$, only the 256 entries of 64 bits are required for cryptography on the group. In comparison with MST3 in Suzuki 2-group, would have 256 entries of 128 bits for $r_i = 2^2$, $s = 64$ and 512 entries for $r_i = 4^2$ , $s = 32$. Also, we should underline that there is a large key data size and the need to compute the inverse element in the finite field. Our future research will be focused on the construction of the cryptosystems on large-order groups.